\documentclass[%
 aip,
 jmp,%
 amsmath,amssymb,
 reprint,%
]{revtex4-1}

\usepackage{graphicx}
\usepackage{dcolumn}
\usepackage{bm}
\usepackage[version=4]{mhchem}
\usepackage{lipsum}
\usepackage{siunitx}
\usepackage{color}

\newcommand{\Liu}{13}

\begin{document}


\title{Magnetoresistance and anomalous Hall effect in micro-ribbons of the magnetic Weyl semimetal \ce{Co3Sn2S2}}

\author{K. Geishendorf}
\affiliation{ 
Leibniz Institute for Solid State and Materials Research (IFW Dresden), Helmholtzstr. 20, 01069 Dresden, Germany
}%
\affiliation{ 
Center for Transport and Devices of Emergent Materials, Technische Universit\"at Dresden, 01062 Dresden, Germany
}%
\author{R. Schlitz}%
\affiliation{ 
Institut f\"ur Festk\"orper- und Materialphysik, Technische Universit\"at Dresden, 01062 Dresden, Germany
}%
\affiliation{ 
Center for Transport and Devices of Emergent Materials, Technische Universit\"at Dresden, 01062 Dresden, Germany
}%
\author{P. Vir}%
\affiliation{ 
Max Planck Institute for Chemical Physics of Solids, 01187 Dresden, Germany
}%
\author{C. Shekhar}%
\affiliation{ 
Max Planck Institute for Chemical Physics of Solids, 01187 Dresden, Germany
}%
\author{C. Felser}%
\affiliation{ 
Max Planck Institute for Chemical Physics of Solids, 01187 Dresden, Germany
}%
\author{K. Nielsch}%
\affiliation{ 
Leibniz Institute for Solid State and Materials Research (IFW Dresden), Helmholtzstr. 20, 01069 Dresden, Germany
}%
\affiliation{ 
Institute of Materials Science, Technische Universit\"at Dresden, 01062 Dresden, Germany
}%
\affiliation{ 
Institute of Applied Physics, Technische Universit\"at Dresden, 01062 Dresden, Germany
}%
\author{S.T.B. Goennenwein}%
\affiliation{ 
Institut f\"ur Festk\"orper- und Materialphysik, Technische Universit\"at Dresden, 01062 Dresden, Germany
}%
\affiliation{ 
Center for Transport and Devices of Emergent Materials, Technische Universit\"at Dresden, 01062 Dresden, Germany
}
\author{A. Thomas}
\affiliation{ 
Leibniz Institute for Solid State and Materials Research (IFW Dresden), Helmholtzstr. 20, 01069 Dresden, Germany
}%
\date{\today}

\begin{abstract}
Magnetic Weyl semimetals exhibit intriguing transport phenomena due to their non-trivial band structure. Recent experiments in bulk crystals of the shandite-type \ce{Co3Sn2S2} have shown that this material system is a magnetic Weyl semimetal. To access the length scales relevant for chiral transport, it is mandatory to fabricate microstructures of this fascinating compound. We therefore have cut micro-ribbons (typical size $0.3~\times~3~\times~$\SI{50}{\micro\meter}$^3$) from \ce{Co3Sn2S2} single crystals using a focused beam of Ga$^{2+}$-ions and investigated the impact of the sample dimensions and possible surface doping on the magnetotransport properties. The large intrinsic anomalous Hall effect observed in the micro ribbons is quantitatively  consistent with the one in bulk samples. Our results show that focused ion beam cutting can be used for nano-patterning single crystalline \ce{Co3Sn2S2}, enabling future transport experiments in complex microstructures of this Weyl semimetal.
\end{abstract}

\maketitle

\section{Introduction}
\noindent Weyl semimetals (WSM) are an interesting class of materials due to their non-trivial topological properties. They posses topologically robust or symmetry protected bulk band crossings in momentum space referred to as Weyl nodes.\cite{Burkov2011} For WSM those nodes always appear in pairs of opposite chirality at different positions in momentum space, connected by Fermi arcs.\cite{Weng2015} The Fermi arcs and Weyl nodes have been probed experimentally in many different materials including \ce{TaAs}, \ce{TaP}, \ce{NbAs}, \ce{NbP}, \ce{WTe2} and \ce{MoTe2} via photoemission spectroscopy and magnetotransport measurements.\cite{Xu2015,Lv2015,Huang2015, Wu2016, Crepaldi2017} A prominent example for the transport signature of Weyl fermions presents the chiral magnetic effect (CME) which gives rise to a positive correction to the magnetoconductance for parallel magnetic and electric field components.\cite{Zhang2016} Weyl nodes can have further implications for transport experiments by acting as sources for Berry curvature.\cite{Wan2011} In other words, the crossing points in momentum space can contribute significantly to the transport in the system if they are located near the Fermi level.\cite{Burkov2015} In turn, the enhanced Berry curvature can give rise to a large intrinsic anomalous Hall effect (AHE), which can be calculated by integrating the Berry curvature of the occupied electronic states.\cite{Fang2003,Xiao2010} One therefore expects a large AHE for magnetic WSM with Weyl nodes close to the Fermi level. Recently shandite type \ce{Co3Sn2S2} was identified as a magnetic WSM. More specifically, \ce{Co3Sn2S2} exhibits a large AHE as well as CME in bulk single crystals.\cite{Liu2018, Wang2018} To explore the transport properties arising from Weyl physics in more detail, it is important to realize microstructures of this fascinating compound. Here we show that \ce{Co3Sn2S2} micro ribbons (typical dimensions of $0.3~\times~3~\times~$\SI{50}{\micro\meter}) with transport properties quantitatively comparable to those of bulk crystals can be fabricated via focused ion beam cutting in combination with conventional lithography techniques. 

\section{Materials and Methods}
\begin{figure*}[hbt]
        \includegraphics[width=0.5\linewidth]{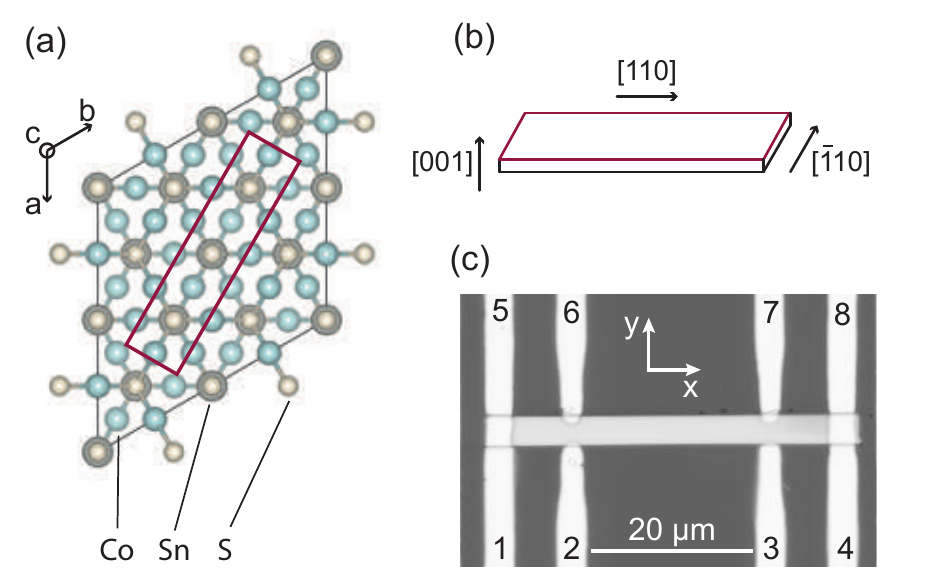}
        \caption{(a) Cross section of the ab-plane highlighting the Kagome lattice formed by the cobalt atoms (light blue). (b) Illustration of the micro-ribbon with crystallographic directions to clarify the orientation with respect to the lattice. (c) Optical micrograph of one of the microdevices fabricated with optical lithography.}
\label{fig:crystalstructure}
\end{figure*}

\noindent\ce{Co3Sn2S2} crystallizes in the shandite crystal structure featuring metallic layers of \ce{Co-Sn} stacked along the c-axis in an ABC fashion where the Co atoms build a corner sharing Kagome lattice (cf. Fig.~\ref{fig:crystalstructure}a).\cite{Kassem2015} Magnetometry measurements on that compound reveal a strong uniaxial anisotropy with an easy axis along the $c$-direction.\cite{Schnelle2013, Kassem2016, Kassem2017} The Kagome lattice with out of plane easy axis makes this material an ideal candidate to study the quantum anomalous Hall effect.\cite{Zyuzin2012}

The single crystals used in this work are grown by a self-flux method with Sn as flux. They grow in thin (hundreds of microns thick) hexagonal shaped flakes with the $c$-direction as the surface normal. We refer to Liu et. al (ref. \Liu) for more details on the growth parameters and bulk properties. We used a focused beam of \ce{Ga^{2+}}-ions to cut the micro-ribbons from the bulk crystals. Focused ion beam (FIB) cutting is a viable tool to fabricate nano- and microstructures based on single crystals.\cite{Moll2010,Moll2016} In our fabrication procedure, a bulk crystal is aligned for the FIB preparation to obtain ribbons with a $c$-direction out of plane. Fig.~\ref{fig:crystalstructure}b illustrates the orientation of the [001], [110] and [$\bar{\text{1}}\text{10}$] within the ribbon. The acceleration voltage for the \ce{Ga^{2+}}-ions is kept constant at \SI{30}{\kilo\volt} during all stages. For FIB preparation the beam aperture size is denoted as current to obtain an energy product when combined with the acceleration voltage. The aperture size is subsequently reduced from \SI{21}{\nano\ampere} for the large scale material removal to \SI{80}{\pico\ampere} for the final shaping in order to reduce the surface deterioration and \ce{Ga^2+} implantation. After FIB preparation the ribbons are lifted out of the single crystal and transferred to a glas substrate for further device preparation. Contacts to the micro-ribbons are fabricated with a standard optical lithography- and lift-off process. To ensure ohmic contacts, the contact areas are cleaned in-situ with \ce{Ar^+}-ion etching before the sputter deposition of \SI{10}{\nano\meter} Cr and \SI{200}{\nano\meter} Au. The samples are subsequently mounted onto chip carriers and inserted in a superconducting magnet cryostat for the transport experiments. In the following, we concentrate on one particular ribbon with dimensions of 0.35~x~4.2~x~\SI{46.1}{\micro\meter}$^3$ which we very systematically studied, but note that the results from this ribbon are consistent with data obtained in further samples.  
Fig.~\ref{fig:crystalstructure}c depicts the device layout used for the magnetotransport measurements. 
The longitudinal resistance $R_{\mathrm{xx}}$ is obtained by applying a current of \SI{100}{\micro\ampere} between contact 1 and 4 while detecting the voltage drop between contact 6 and 7. The transverse (Hall) voltage drop $V_\mathrm{y}$ is simultaneously detected between contacts 2 and 6 while sweeping the magnetic field ($\mathrm{\mu}_0H_\mathrm{z}$) along the $z$-direction. For calculating the resistivities, the ribbon thickness is determined to \SI{350}{\nano\meter} by scanning electron microscopy.

\section{Results and Discussion}
\begin{figure*}[hbt]
\includegraphics[width=0.5\linewidth]{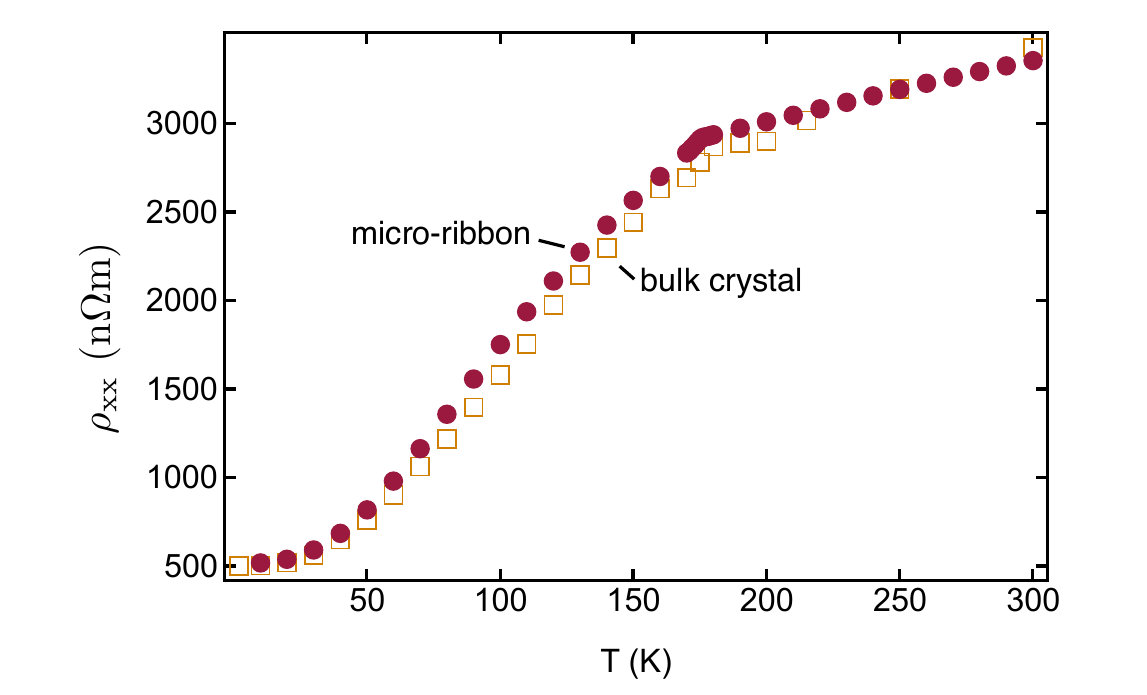}
\caption{Exemplary longitudinal resistivity $\rho_{\mathrm{xx}}$ of a 0.35~x~4.2~x~\SI{46.1}{\micro\meter}$^3$ \ce{Co3Sn2S2} micro-ribbon in the temperature range from 10 to \SI{300}{\kelvin} (solid circles) compared to the values reported for bulk crystals (open rectangles, taken from ref. \Liu).}
\label{fig:RT_device}
\end{figure*}

\ce{Co3Sn2S2} features a positive temperature coefficient of the resistivity for the entire temperature range from \SI{300}{\kelvin} to \SI{10}{K} studied here. Indeed, as evident from Fig.~\ref{fig:RT_device}, the resistivity decreases from \SI{3.2}{\micro\ohm\meter} to \SI{0.5}{\micro\ohm\meter}. The resistivity is proportional to the temperature for $T$\textgreater$T_{\mathrm{c}}$ suggesting that the resistivity is governed by electron-phonon interactions in that region.\cite{Kasuya1956} A pronounced change of slope is evident at $T$=\SI{174}{\kelvin} which is consistent with the Curie temperature of this compound. \cite{Schnelle2013,Kassem2016, Kubodera2006, Corps2015} The steeper $\mathrm{d}\rho /\mathrm{d}T$ for $T$\textless$T_\mathrm{c}$ is related to the spin-disorder contribution to the resistivity which is frozen out below $T_\mathrm{c}$ as the moments align. \cite{Kasuya1956} For bulk samples a residual resistance ratio (RRR) of 6.8 was found, which is closely comparable to the RRR of 6.5 which we obtain in our micro-ribbons.\cite{Liu2018}

\begin{figure*}[hbt]
\includegraphics[width=\linewidth]{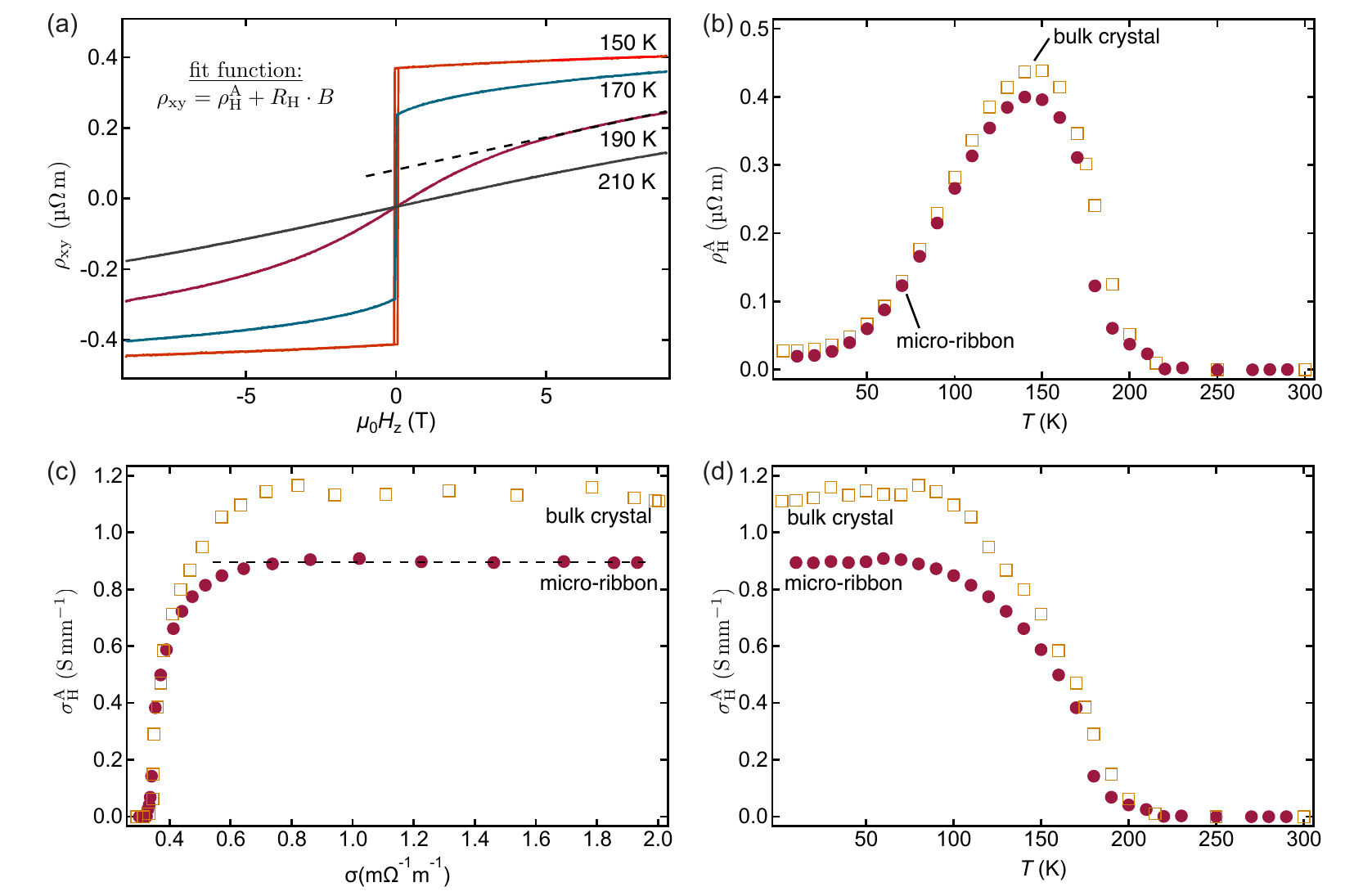}
\caption{(a) Exemplary Hall resistivity  as a function of magnetic field, recorded for three temperatures around $T_\mathrm{c}$. The linear fit to the high field part ($>$\SI{5}{\tesla}) is used to extract the AHE. (b) Anomalous Hall resistivity as a function of temperature for the micro ribbon (solid circles) compared to bulk values (open squares, taken from Ref. \Liu). (c) The AHC as a function of the conductance, showing a wide plateau as it is characteristic for intrinsic AHE. (d) The AHC is very weakly temperature dependent for $T$\textless\SI{100}{\kelvin},in good agreement with the intrinsic AHE contribution predicted from band structure calculations.}
\label{fig:Hall}
\end{figure*}

We now turn to the Hall response. The bulk crystals feature a giant AHE of up to \SI{0.44}{\micro\ohm\meter} being identified as contribution of the Weyl nodes to the Berry curvature.\cite{Liu2018} It has been pointed out that the intrinsic AHE is sensitive to the exact position of the Weyl nodes with respect to the Fermi level because only Weyl nodes close to the Fermi level strongly contribute to the transport properties. The dependence of the anomalous Hall conductance (AHC) on the Fermi level position was calculated by Liu and coworkers.\cite{Liu2018} A sharp peak of up to \SI{1.1}{\siemens\per\milli\meter} is expected for the pure \ce{Co3Sn2S2} compound. From these calculations one would expect a strong decrease of the AHC for an upwards shift of the Fermi energy whereas it is rather robust against lowering the Fermi level. In previous investigations of magnetotransport in FIB fabricated micro-ribbons it has been found that the implantation of \ce{Ga^{2+}}-ions leads to a noticeable surface doping effect.\cite{Niemann2017} Thus, to gauge possible effects of the sample preparation on the location of the Weyl points, we compare the AHC of the micro-ribbon with the values reported for bulk crystals. Exemplary Hall measurements for temperatures in the vicinity of $T_\mathrm{c}$ are depicted in Fig. \ref{fig:Hall}a. The anomalous Hall resistivity is deduced by linear fitting the high field part ($\mu_0H_{\mathrm{z}}>$\SI{5}{\tesla}) and determining the intercept ($\rho_{\mathrm{H}}^{\mathrm{A}}$) of the linear fit with the ordinate. This corresponds to the transverse voltage response arising from the saturation magnetization without any external magnetic field. \cite{Nagosa2010} To correct possible offsets, arising e.g. due to non-ideal contact placement, $\rho_{\mathrm{H}}^{\mathrm{A}}$ is antisymmetrized following ($\rho_{\mathrm{H}}^{\mathrm{A}}=\mathrm{(}\rho_{\mathrm{H}}^{\mathrm{A}}(B)-\rho_{\mathrm{H}}^{\mathrm{A}}(-B)\mathrm{)/}2$). Above \SI{220}{\kelvin}, the transverse resistivity ($\rho_{\mathrm{xy}} = -V_{\mathrm{y}}/I_{\mathrm{x}} \cdot t$) depends linearly on the field as expected for the ordinary Hall effect not shown here.\cite{Nagosa2010} For lower temperatures an nonlinear Hall contribution emerges, resulting in s-shaped Hall traces and thus a finite $\rho_{\mathrm{H}}^{\mathrm{A}}$. The extracted $\rho_{\mathrm{H}}^{\mathrm{A}}$ is plotted in Fig. \ref{fig:Hall}b as a function of temperature where it displays a steep increase below $T_\mathrm{c}$ with a maximum of \SI{0.4}{\micro\ohm\meter} between \SIlist{140;150}{\kelvin} as it was already shown for bulk crystals (cf. open squares in Fig. \ref{fig:Hall}). Interestingly, \ce{Co3Sn2S2} shows a finite AHE above $T_\mathrm{c}$ up to about \SI{220}{\kelvin}. To evaluate the microscopic mechanisms involved in the AHE, it has proven helpful to study the evolution of $\sigma_{\mathrm{H}}^{\mathrm{A}}$ with $\sigma_{\mathrm{xx}}$ (cf. Fig.~\ref{fig:Hall}c). Since the mechanisms exhibit a distinct scaling behavior with $\sigma_{\mathrm{xx}}$ such a study allows to probe the involved effects.\cite{Nagosa2010} \ce{Co3Sn2S2} shows $\sigma_{\mathrm{H}}^{\mathrm{A}} \propto \sigma^0$ for a conductance between \SIlist{0.7;1.9}{\per\milli\ohm\per\meter}.

Such a plateau is characteristic for the intrinsic AHE contribution or the side-jump mechanism. According to the calculations in ref.~\Liu, the intrinsic contribution should be in the order of \SI{1.1}{\siemens\per\milli\meter} in good quantitative agreement with both the bulk and micro-ribbon results. It is further characteristic for the intrinsic AHE contribution that its magnitude is independent of temperature (for $T$\textless$T_\mathrm{c}$) but only depends on the magnetization. This behaviour is observed for \ce{Co3Sn2S2} and becomes evident by comparing Fig.~\ref{fig:Hall}d to $M$(H) curves reported previously.\cite{Sakai2013,Corps2015} Taken together, Fig.~\ref{fig:Hall}b-d shows that the micro-ribbon and the bulk crystals show quantitatively similar transport properties.

\begin{figure*}[hbt]
\includegraphics[width=\linewidth]{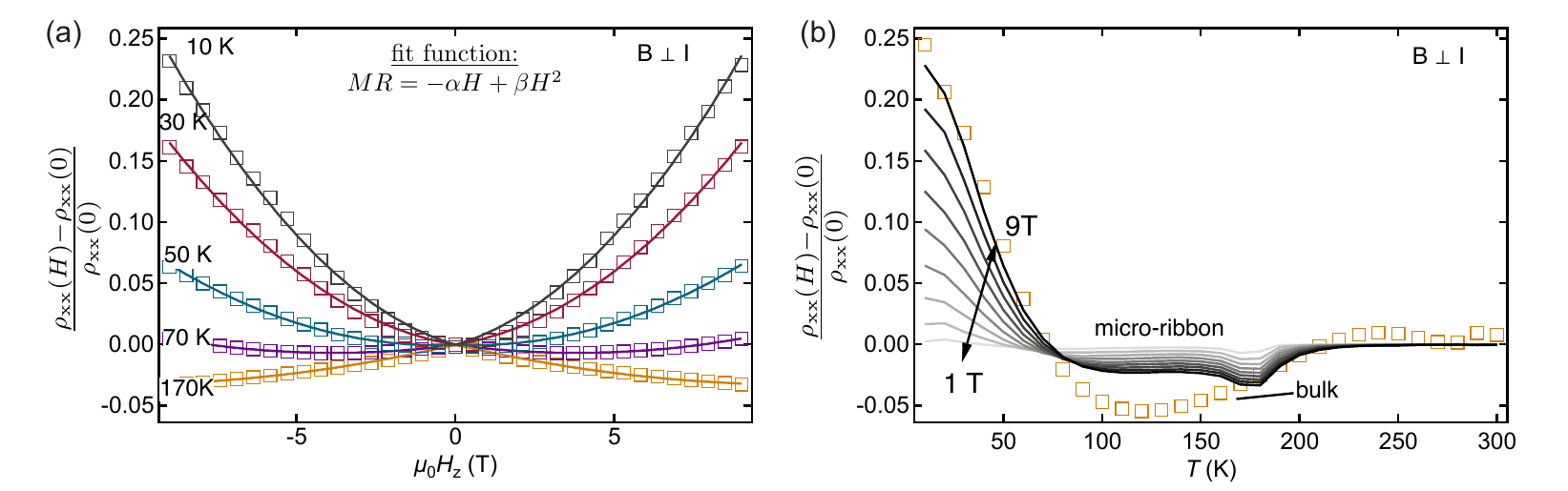}
\caption{(a) Magneto-resistance (MR) recorded with the magnetic field oriented in out of plane orientation ($B \perp E$). The open squares are the measurement data and the solid lines are fits to a model with contributions from the suppression of spin fluctuations and the Lorentz deflection of charge carriers by an external magnetic field. (b) MR as function of temperature plotted for different external magnetic fields with magnitudes between \SIlist{1;9}{\tesla}.  The data for the bulk crystal is taken from Ref.~\Liu}
\label{fig:MR}
\end{figure*}

We now turn to evolution of the magnetoresistance (MR) with magnetic field and temperature. The MR is obtained from the longitudinal resistance $\rho_{\mathrm{xx}}$ during sweeps of the magnetic field along the $z$-direction (out of plane). Figure~\ref{fig:MR}a depicts exemplary MR measurements (open squares), where the MR is derived as ($\rho_{\mathrm{xx}}(H)~-~\rho_{\mathrm{xx}}(0))/\rho_{\mathrm{xx}}(0)$. Extracting the MR magnitude for different field strength as a function of temperature (cf. Fig.~\ref{fig:MR}b) gives further insight into the different contributions to the MR. The open squares in Fig.~\ref{fig:MR}b show the data for the bulk crystal extracted from ref.~\Liu~in the same manner for \SI{9}{\tesla}. \ce{Co3Sn2S2} exhibits a small negative MR (\textless 0.5\%) in the paramagnetic state which strongly increases in vicinity of $T_\mathrm{c}$ to a magnitude of 3.3\% at \SI{9}{\tesla}. The compound displays a constant negative MR of 2.2\% at \SI{9}{\tesla} between \SIlist{110;150}{\kelvin}. Below \SI{110}{\kelvin} a crossover to positive non-saturating MR is observed reaching a maximum of 22.8\% at \SI{9}{\tesla}. This behavior can be explained by two different contributions, first the suppression of spin-fluctuations and second the Lorentz deflection of charge carries due to an external magnetic field. In the ferromagnetic state a linear negative MR is expected from the suppression of spin fluctuations. \cite{Ueda1976} A maximum of the negative MR at $T_\mathrm{c}$ is also observed in other ferromagnetic compounds due to the significant suppression of critical fluctuations near the ordering temperature.\cite{Masuda1977} This behavior is only apparent in the microstructures presented here, whereas no sharp maximum of the negative MR around $T_\mathrm{c}$ can be observed in the bulk crystals (cf. open squares in Fig.~\ref{fig:MR}b). The second contribution from the Lorentz deflection is proportional to the charge carrier mobility. Semimetals typically exhibit high mobilities at low temperatures. \cite{Ali2014, Liang2015, Kumar2017} Therefore, when decreasing the temperature the Lorentz deflection becomes the governing MR mechanism and this deflection can be captured with Kohler's law where the MR is proportional to $H^2$. A simple model with a negative term proportional to H and a positive term scaling with $H^2$ is consistent with the measured MR in \ce{Co3Sn2S2} (cf. solid lines in Fig~\ref{fig:MR}).

\section{Conclusions}

It is not straight forward for microstructures of exotic materials, such as the recently discovered magnetic WSM \ce{Co3Sn2S2}, to resemble the fascinating properties observed in bulk samples. We have therefore, cut several micro-ribbons from \ce{Co3Sn2S2} single crystals using FIB and investigated their transport properties by complementary magnetoresistance- and Hall effect measurements. The magnetoresistance (for field perpendicular to the current flow direction) can be consistently explained by contributions from the suppression of spin-fluctuations and Lorentz deflection of high mobility charge carriers in semimetals. More importantly, the quantitative agreement between the AHE observed in micro-ribbons and the one reported for bulk samples makes clear that the intriguing Weyl properties are not affected by the FIB microstructuring process. In other words, our data consistently shows that reducing the sample dimensions down to hundreds of nanometers does not change the salient features observed in the transport properties. In particular, the implantation of \ce{Ga^{2+}}-ions (which inevitably occurs during the FIB process) does not significantly alter the properties of \ce{Co3Sn2S2}. We thus have shown that FIB cutting can be used for nano-patterning single crystalline \ce{Co3Sn2S2}, enabling future experiments in complex nano- and microstructures. This will allow further study of the implications of Weyl node contributions to the charge transport.

\section{Acknowledegments}
The authors gratefully acknowledge the FIB preparation by Almut P\"ohl and Tina Sturm. Financial support via DFG SFB 1143 (project B05, C05 and C08) is gratefully acknowledged. 

\pagebreak

\clearpage

\bibliography{references.bib}

\end{document}